# Integrating upstream and downstream reciprocity stabilizes cooperator-defector coexistence in others-only public goods games

**Tatsuya Sasaki[1*], Satoshi Uchida[2,3], Isamu Okada[4], Hitoshi Yamamoto[5], Yutaka Nakai[6]**

[1] Department of Community Development, Koriyama Women's College, Koriyama, Japan
[2] Research Center for Ethi-Culture Studies, RINRI Institute, Tokyo, Japan
[3] High-tech Research Centre, Kokushikan University, Tokyo, Japan
[4] Faculty of Business Administration, Soka University, Tokyo, Japan
[5] Faculty of Business Administration, Rissho University, Tokyo, Japan
[6] Research Institute for Socionetwork Strategies, Kansai University, Osaka, Japan

*** Correspondence:**
Tatsuya Sasaki
t.sasaki@koriyama-kgc.ac.jp

**Abstract**

Introduction: Human cooperation persists among strangers in large, well-mixed populations despite theoretical predictions of difficulties, leaving a fundamental evolutionary puzzle. While upstream (pay-it-forward: helping others because you were helped) and downstream (rewarding-reputation: helping those with good reputations) indirect reciprocity have been independently considered as solutions, their joint dynamics in multiplayer contexts remain unexplored.

Methods: We study public goods games (PGGs) without self-return (often called "others-only" PGGs) with benefit $b$ and cost $c$ and analyze evolutionary dynamics for three strategies: unconditional cooperation (X), unconditional defection (Y), and an integrated reciprocal strategy (Z) combining unconditional forwarding with reputation-based discrimination.

Results: We show that integrating upstream and downstream reciprocity can yield a globally asymptotically stable mixed equilibrium of unconditional defectors and integrated reciprocators when the benefit-to-cost ratio exceeds a threshold ($b/c > 2$) in the absence of complexity costs. We analytically derive a critical threshold for complexity costs. If cognitive demands exceed this threshold, the stable equilibrium disappears via a saddle-node bifurcation. Otherwise, within the stable regime, complexity costs counterintuitively stabilize the equilibrium by preventing not only unconditional cooperators (viewed as second-order freeloaders) but also alternative conditional strategies from invading.

Discussion: Rather than requiring uniformity, our model reveals one pathway to stable cooperation through strategic diversity—defectors serve as "evolutionary shields" preventing system collapse while integrated reciprocators flexibly combine open and discriminative responses. This framework demonstrates how pay-it-forward broadcasting and reputation systems can jointly maintain social polymorphism including cooperation despite cognitive limitations and group size challenges, offering a potential evolutionary foundation for behavioral diversity in human societies.

# 1 Introduction

Reciprocal cooperation is a cornerstone of sustainable societies (Trivers, 1971). Yet empirical regularities show that such cooperation, which thrives in small-scale, face-to-face environments, erodes as group size increases (Shinada and Yamagishi, 2008). As communities expand, opportunities for mutual aid with the same people decrease, so a generalized reciprocity system is expected to function, in which when individuals provide resources or assistance to others, they do not receive a direct benefit but rather receive a benefit from another member of the community at a later date (Takahashi, 2000; Mashima and Takahashi, 2008).

Such indirect forms of reciprocity, have been proposed in two distinct ways: (i) upstream reciprocity, often described as "pay-it-forward," and (ii) downstream reciprocity, also known as "rewarding reputation" (Boyd and Richerson, 1989; Nowak and Sigmund, 2005; Baker and Bulkley, 2014; Watanabe et al., 2014). Although each mechanism can, in isolation, elicit helping among strangers, both become fragile under freeloading, participation costs, or complexity costs of tracking both emotional and reputational information (Brandt and Sigmund, 2006; Peña et al., 2011; Suzuki and Kimura, 2013; Yamamoto et al., 2024; see also Imhof et al., 2005). Moreover, the empirical reality that cooperators and freeloaders persistently coexist, especially in larger groups, hints that mixed behavioral strategies may be adaptive rather than problematic. Understanding how integration of these mechanisms could maintain such diversity under harsh conditions, therefore, remains an open problem in the evolutionary theory of cooperation.

While these mechanisms are often conceptualized as dyadic chains, in group settings, they frequently manifest as "broadcasting" or one-to-many propagation rather than strictly linear interactions. For instance, in digital communities, sharing knowledge on forums (e.g., StackOverflow) or posting exclusive invitation codes incurs a personal cost but benefits multiple observers simultaneously. Similarly, in organizational contexts, bringing souvenirs or refreshments for a team functions as a "bundle of mutual gifts" to the group, potentially triggering upstream reciprocity that propagates across departments. Anthropologically, this structure mirrors food sharing in hunter-gatherer societies, where large game is distributed to the entire group. These real-world dynamics align with the structure of public goods games without self-return (others-only PGGs), where a donor's action benefits all other group members simultaneously.

To understand how these mechanisms function and potentially interact, it helps to examine their distinct operational logics in everyday contexts.

(i) Upstream reciprocity ("pay-it-forward") operates through emotional contagion—receiving help triggers gratitude or indebtedness that motivates helping others, propagating generosity that can flow forward even through populations (Brandt and Sigmund, 2006; Nowak and Roch, 2007; Daimon and Atsumi, 2021; Obayashi et al., 2023). For instance, researcher Alice receives language assistance from a colleague; gratitude motivates Alice to assist a third party with a task. In *upstream* indirect reciprocity, the focal recipient helps a third party rather than the original donor.

(ii) Downstream reciprocity ("rewarding reputation"), by contrast, often operates through strategic calculation—individuals selectively help those with good reputations to build their own standing, creating incentive structures for cooperation (Alexander, 1987; Nowak and Sigmund, 1998a,b). For instance, researcher Bob chooses to help a colleague already known for generosity, aiming thereby to enhance Bob's own standing. Here, the decision is guided by publicly shared reputation rather than by emotion. In *downstream* indirect reciprocity, the focal donor receives help from a third party rather than the original recipient.

Field evidence indicates that both pathways can promote cooperation and, importantly, can reinforce one another when they co-occur (Baker and Bulkley, 2014; Simpson et al., 2018). Whether such hybrid dynamics can survive in larger groups, where individuals must rely on such costly cognitive systems as tracking both emotions and reputations, is still unclear.

Most formal models continue to treat upstream and downstream reciprocity in isolation and to restrict attention to dyadic interactions (Boyd and Richerson, 1989; Brandt and Sigmund, 2006); in addition, multi-player studies typically presuppose cost-free information processing and purely reputation-based reaction rules (Suzuki and Akiyama, 2005, 2007, 2008; see also Wei et al., 2024). Consequently, we lack (i) an analytic account of how the two indirect reciprocity modes interact in larger groups and (ii) a quantitative assessment of the complexity cost incurred when individuals track both personal affect and third-party images. These omissions leave unanswered the question of whether an integrated strategy can outperform simpler indiscriminators once group size and cognitive load are explicitly varied.

Here, we present a standard evolutionary-game model that integrates pay-it-forward-driven generosity with reputation-based partner selection in a one-shot linear public goods game without self-return, often referred to as the others-only PGG (Sigmund, 2010; van Veelen, 2020). Using replicator dynamics (Hofbauer and Sigmund, 1998), we (i) derive closed-form conditions under which the integrated strategy coexists with selfish competitors, (ii) demonstrate how the equilibrium frequency of integrated reciprocators declines with group size, and (iii) show that a modest complexity cost can stabilize polymorphism rather than precipitate a collapse of cooperation. These results not only extend previous dyadic analyses (Sasaki et al., 2024) but also offer a paradigm shift from "solving cooperation" to "maintaining diversity."

The remainder of the paper is organized as follows. Section 2 introduces the others-only PGG and formalizes the action rules and the three evolutionary strategies (unconditional cooperation, unconditional defection, and integrated reciprocation). Section 3 specifies the payoff structure including the complexity cost, derives the replicator dynamics, and characterizes the equilibrium frequency of integrated reciprocators. We also examine how increases in group size and complexity cost reshape the stability landscape. Section 4 concludes the study and outlines promising directions for theoretical extensions.

## 2 Materials and methods

### 2.1 Others-only public goods games and action rules

We consider an infinitely large, well-mixed population. In each interaction, $N \geq 2$ players are drawn at random and interact for a single round (one shot). Within the group, all $N$ players simultaneously act as donors in the others-only PGG: A donor can choose whether to cooperate or not; if cooperating, a donor pays a cost $c > 0$ to confer a benefit $b > c$. The contributed benefit is then equally shared (i) among all of the $N - 1$ co-players in the case of universal cooperation (UC) or (ii) among only the recipients whom the donor selects from the $N - 1$ co-players in the case of conditional cooperation (CC). Universal non-cooperation (UD) does not affect either the donor or the co-players. After the round, the group dissolves and a new group of $N$ players is randomly formed. Groups are reshuffled independently across rounds (memoryless matching).

We note that strictly speaking, reciprocity implies a sequential interaction. However, following standard approaches in evolutionary game theory (Sigmund, 2010), we assume a simultaneous, symmetric setting to simplify the payoff formulas. This structure is mathematically equivalent to a linear PGG without self-return (Sigmund, 2010; van Veelen, 2020). For $N = 2$, this formulation naturally yields the standard pairwise giving game (or donation game).

The case (i) is a straightforward expansion of the pairwise (or 1-to-1) giving game (Nowak and Sigmund, 1998a,b; Panchanathan and Boyd, 2003) to its $N$-player (or 1-to-$N-1$) version. In others-only PGGs with UC and UD, the focal player who plays either UC or UD with $N-1$ co-players yields the payoffs, as follows, respectively:

$$P(\mathrm{UC}, k) = \frac{kb}{N-1} - c,$$

and

$$P(\mathrm{UD}, k) = \frac{kb}{N-1},$$

where k denotes the number of UC-players among $N-1$ co-players. This indicates that the payoff for each player (whether UC or UD) in the group increases with the number of UC players, $k$, and also that switching from UC to UD leads to an improvement in the payoff regardless of the choices of the other players. UD dominates UC.

## 2.2 Evolutionary strategies

This study investigates the evolutionary dynamics of three strategies: unconditional cooperation (X), unconditional defection (Y), and conditional cooperation (Z). For conditional cooperators, the study particularly considers integrated reciprocators, which are defined as unifying upstream and downstream reciprocity (Fig. 1).

Sasaki et al. first proposed a model of integrated reciprocity in the giving game (others-only PGG with $N = 2$) (Sasaki et al., 2024). In the two-player model, integrated reciprocators are characterized by giving benefits $b$ to whoever their co-players are if they received any benefits in the previous round. Otherwise, they offer benefits to co-players who have a Good image.

This study extends the pairwise model to $N$-player interactions. First, integrated reciprocators (Z) who received benefits in the previous round will universally cooperate with all the other co-players. The benefits offered are then equally shared among all of the $N-1$ co-players, who thus each receive $b/(N-1)$, as in playing UC (Table 1). This conditional play of UC is the so-called "pay-it-forward."

Second, integrated reciprocators (Z) who received no benefit in the previous round will conditionally cooperate only with the other players who have a Good image. The benefits offered are equally shared among the Good co-players. In this game (others-only PGG), the focal donor never rewards herself, and hence, each Good recipient obtains $b/(\text{the number of their Good co-players} - 1)$ (Table 1).

The above conditional play of CC is referred to as “rewarding reputation.” We note that integrated reciprocators willing to reward Good co-players are not allowed to reward themselves. Thus, our multiplayer games provide no self-return for cooperative players.

Unconditional cooperators (X) are defined as always unconditionally playing UC. The benefit provided by an X-player is equally distributed to all of the $N - 1$ co-players in the group. In contrast, unconditional defectors (Y) are defined as always unconditionally playing UD.

### 2.3 Population dynamics and image assessment

The evolutionary dynamics of the three strategies take place in the state space $S_3 = \{(x, y, z) | x + y + z = 1; x, y, z \geq 0\}$. The three homogeneous states in which 100% of the population are X-players ($x = 1$), Y-players ($y = 1$), and Z-players ($z = 1$) correspond to three vertices of the simplex $S_3$ (which we denote by X, Y, and Z, respectively). These are trivial fixed points for the replicator system. There are no other fixed points on the edge X-Y along which the evolution is unidirectional from X to Y.

We employ a strict assessment rule, which functions as a “dual-verification” social norm, to distinguish integrated reciprocators from other strategies. An individual is assessed as “Good” if and only if they satisfy both of the following action-based criteria (an AND condition): (i) Upstream Check: cooperating with co-players if the individual received help in the previous round, and (ii) Downstream Check: cooperating selectively with “Good” co-players (and defecting against “Bad” ones) if the individual did not receive help. Failing either condition results in a Bad image. In a mixed population containing unconditional defectors (Y), this rule allows for the discrimination of unconditional cooperators (X). Since X-players indiscriminately help Bad Y-players, they fail the Downstream Check and are consequently assessed as Bad. In contrast, Z-players correctly defect against Y-players, passing both checks and maintaining a Good image.

This strict assessment rule provides the microscopic foundation for the complexity cost $d$. Unconditional strategies, X and Y, are cognitively “blind” in the sense that they act without monitoring previous interactions or social standings. Consequently, we assume they incur no information-processing costs. In contrast, the integrated strategy Z requires tracking both personal history (upstream trigger) and the reputation of others (downstream verification) to execute the dual-verification norm. We define $d$ as the specific cognitive or physiological cost associated with this additional information processing. While the magnitude of $d$ may be small, it is strictly positive ($d > 0$), physically distinguishing the conditional strategy Z from the unconditional strategies X and Y.

Regarding the initial state of the dynamics ($t = 0$), we follow standard practices in indirect reciprocity literature (Panchanathan and Boyd, 2003; Ohtsuki and Iwasa, 2007). We assume that all players initially possess a Good image and that no player has received help in a “previous” round. Under these conditions, in the first round, Z-players follow their downstream reciprocity logic: Since they have not received help and all opponents are initially Good, they cooperate. Y-players defect. Consequently, immediately after the first round, Y-players are assessed as Bad (having failed to cooperate), while Z-players maintain their Good image. Following this transient phase, the system quickly converges to the stationary state where the probability of receiving help, $u$, is endogenously determined by the population frequency $z$.

## 3 Results

### 3.1 Expected payoffs

For simplicity, let us first focus on the one-dimensional replicator dynamics for integrated reciprocators (Z) and unconditional defectors (Y), in which the subpopulation of unconditional defectors is a complement of that of integrated reciprocators, thus $y = 1 - z$.

Let $u$ denote the probability that a focal Z-player received benefits in the previous round (from either pay-it-forward or rewarding). Under memoryless matching, the indicator "helped in the previous round" for a focal Z-player is independent of the current opponents; hence we model Z-player's action as a mixture: With probability $u$ it plays pay-it-forward (UC), and with probability $1 - u$ it plays rewarding reputation (CC). Additionally, we denote $N_Z$ as the number of Z-players in the group. Z-players now coincides with Good-players, and thus $N_Z$, the number of Good-players in the group. We will derive the expected payoff for Z-players in a group with $N_Z$ Z-players.

The expected payoff $F_{Z,1}(N_Z)$ for the focal Z-player, which is issued from pay-it-forward, is as follows:

$$
\begin{aligned}
&F_{Z,1}(N_Z)\\
&= \frac{1}{N-1}\sum_{m=0}^{N_Z-1}\binom{N_Z-1}{m}u^m(1-u)^{(N_Z-1)-m}mb - uc\\
&= \frac{(N_Z-1)ub}{N-1}\sum_{m=0}^{N_Z-1}\binom{N_Z-2}{m-1}u^{m-1}(1-u)^{(N_Z-2)-(m-1)} - uc\\
&= \frac{(N_Z-1)ub}{N-1}\sum_{m'=0}^{N_Z-2}\binom{N_Z-2}{m'}u^{m'}(1-u)^{(N_Z-2)-m'} - uc\\
&= \frac{(N_Z-1)ub}{N-1} - uc,
\end{aligned}
\tag{1-1}
$$

in which the first term expresses the benefits obtained from other pay-it-forward Z-players in the group, and the second term, the costs to the focal pay-it-forward Z-player.

Moreover, the expected payoff $F_{Z,2}(N_Z)$ for the focal Z-player, which is issued from rewarding-reputation, is as follows: for $N_Z \geq 2$,

$$
\begin{aligned}
&F_{Z,2}(N_Z)\\
&= \frac{1}{N_Z-1}\sum_{m=0}^{N_Z-1}\binom{N_Z-1}{m}u^m(1-u)^{(N_Z-1)-m}[(N_Z-1)-m]b - (1-u)c
\end{aligned}
$$

$$= \frac{(N_Z - 1)(1-u)b}{N_Z - 1} \sum_{m=0}^{N_Z-2} \binom{N_Z - 2}{m} u^m (1-u)^{(N_Z-2)-m} - (1-u)c$$

$$= (1-u)b - (1-u)c,$$

(1-2)

in which the first and second terms give the benefits obtained from other rewarding-reputation Z-players in the group and the costs to the focal rewarding-reputation Z-player, respectively.

Otherwise, for $N_Z = 1$, which means that the focal player is the only one conditional reciprocator, we have,

$$F_{Z,2}(1) = 0,$$

(1-3)

that is, no conditional reciprocator in co-players leads to no rewarding.

Using the above equations, the expected payoff for the conditional reciprocator with the other $(N_Z - 1)$-conditional reciprocators (Z) in the group, $F_Z(N_Z)$, is given by

$$F_Z(N_Z) = F_{Z,1}(N_Z) + [1 - (1-z)^{N-1}] F_{Z,2}(N_Z) - d,$$

(1-4)

in which $d (> 0)$ denotes the complexity costs and the square bracket term describes the probability that there exists at least one Z-player among the co-players in the group.

In contrast, Y-players never receive benefits from rewarding-reputation by Z-players. Y-players can receive benefits if and only if Z-players in the same group received the benefits in the previous round with probability $u$, in which case Z-players are willing to cooperate with all players. The expected payoff for Y-players in a group with $N_Z$ Z-players is

$$F_Y(N_Z)$$

$$= \frac{1}{N-1} \sum_{m=0}^{N_Z} \binom{N_Z}{m} u^m (1-u)^{N_Z-m} mb$$

$$= \frac{N_Z ub}{N-1} \sum_{m=1}^{N_Z} \binom{N_Z - 1}{m-1} u^{m-1} (1-u)^{(N_Z-1)-(m-1)}$$

$$= \frac{N_Z ub}{N-1} \sum_{m'=0}^{N_Z} \binom{N_Z - 1}{m'} u^{m'} (1-u)^{(N_Z-1)-m'}$$

$$= \frac{N_Z ub}{N-1}. \tag{2}$$

We calculate the probability, $u$, that Z-players receive benefits in a round. First, the case must hold that if a Z-player exists among co-players, the focal Z-player will receive benefits. This occurs with the probability of $1-(1-z)^{N-1}$, in which case the co-Z-player is willing to do either pay-it-forward or rewarding-reputation. We note that by a mean-field, stationarity assumption we identify last-round $z$ with the current $z$. Considering that the probabilities of doing pay-it-forward and rewarding-reputation are $u$ and $1-u$, respectively, the probability for Z-players to receive benefits is given by

$$u = [1-(1-z)^{N-1}][u+(1-u)] = 1-(1-z)^{N-1}. \tag{3}$$

### 3.2 Replicator dynamics

Let $F_Z$ and $F_Y$ denote the expected payoffs for Z- and Y-players, respectively, calculated over all possible group compositions. Based on Eqs. (1) to (3), these are defined as follows:

$$F_Z$$

$$= \sum_{N_Z=1}^{N} \binom{N-1}{N_Z-1} z^{N_Z-1}(1-z)^{(N-1)-(N_Z-1)} \frac{(N_Z-1)ub}{N-1}$$
$$- uc + [1-(1-z)^{N-1}](1-u)(b-c) - d,$$

By setting $S = N_Z - 1$, we have

$$= \frac{ub}{N-1}\sum_{S=0}^{N-1} S\binom{N-1}{S} z^{S}(1-z)^{(N-1)-S}$$
$$- uc + [1-(1-z)^{N-1}](1-u)(b-c) - d$$

$$= \frac{ub}{N-1}(N-1)z - uc + [1-(1-z)^{N-1}](1-u)(b-c) - d$$

$$= zub - uc + [1-(1-z)^{N-1}](1-u)(b-c) - d, \tag{4}$$

and

$$F_Y$$

$$= \sum_{N_Z=0}^{N-1} \binom{N-1}{N_Z} z^{N_Z}(1-z)^{(N-1)-N_Z} \frac{N_Z ub}{N-1}$$

by setting $S = N_Z - 1$, we have

$$= \frac{ub}{N-1} \sum_{S=0}^{N-2} \binom{N-2}{S} (N-1) z^{S+1} (1-z)^{(N-2)-S}$$

$$= \frac{ub}{N-1} (N-1) z \sum_{S=0}^{N-2} \binom{N-2}{S} (N-1) z^{S} (1-z)^{(N-2)-S}$$

$$= \frac{ub}{N-1} (N-1) z$$

$$= zub.$$

(5)

These lead to

$$F_Z - F_Y$$

$$= -uc + [1 - (1-z)^{N-1}](1-u)(b-c) - d.$$

(6)

And then, substituting Eq. (3) into $u$, we have

$$F_Z - F_Y$$

$$= [1 - (1-z)^{N-1}][-c + (1-z)^{N-1}(b-c)] - d$$

$$= z \left( \sum_{k=0}^{N-2} (1-z)^k \right) [-c + (1-z)^{N-1}(b-c)] - d.$$

(7)

The replicator dynamics of integrated reciprocators (Z) and unconditional defectors (Y) are thus given by

$$\dot{z} = z(1-z)(F_Z - F_Y).$$

(8)

Equation (7) takes zero for $z = 0, 1$. Its left bracket term is positive for $z < 1$, and its right bracket strictly monotonically decreases with $z$ because $(1-z)^{N-1}$ also does, leading to (i) a unique root for $d = 0$ and (ii) a couple of roots for sufficiently modest $d$, in the open interval $(0,1)$.

### 3.2.1 Case of no complexity cost $d = 0$

We first investigate the evolutionary dynamics with no complexity cost.

**Proposition.** In the others-only PGG with integrated reciprocity and $d = 0$, an interior Y–Z coexistence exists if and only if $b/c > 2$, given by

$$z_0 = 1 - \left(\frac{c}{b-c}\right)^{\frac{1}{N-1}}. \tag{9}$$

Setting P: $z = z_0$, P is asymptotically stable and strictly decreases with $N$.

*Proof.* Since $(1-z)^{N-1}$ is strictly decreasing for $z$, the replicator equation in Eq. (8) has an interior solution iff $0 < c/(b-c) < 1$, i.e., $b/c > 2$. Uniqueness follows from the monotonicity of $(1-z)^{N-1}$. For stability, note $F_Z - F_Y > 0$ near $z = 0$ when $b > 2c$, and $F_Z - F_Y < 0$ near $z = 1$; hence the interior root is asymptotically stable. Monotonicity in $N$ follows because for $\alpha \in (0,1)$, $\alpha^{1/(N-1)}$ increases with $N$. ■

Subsequently, it follows that, when we have $b/c > 2$ , the attractor P enters the state space from $z = 0$, and also that as the benefit-to-cost ratio $b/c$ increases, $z_0$ monotonically converges to 1 and P moves infinitely close to the other side of the state space.

In the specific case with $N = 2$, Eqs. (8) and (9) turn to

$$F_Z - F_Y = z[-c + (1-z)(b-c)], \tag{10}$$

and

$$z_0 = \frac{b-2c}{b-c}, \tag{11}$$

respectively. This is consistent with the results by Sasaki et al. (2024) in the pairwise giving game. As shown in Fig. 2, it follows from Eq. (9) that as the group size increases, the attractor P moves toward the corner, which represents the 100% state of unconditional defection, thereby decreasing the frequency of cooperation.

### 3.2.2 Case of complexity cost $d > 0$

**Remark.** For sufficiently small $d > 0$, two edge-interior equilibria $\mathrm{P_Y}(d)$ (unstable, near node Y) and $\mathrm{P_Z}(d)$ (stable, near the $d = 0$ root) bifurcate; the coexistence persists for $b/c > 2$ until $d$ reaches a finite saddle-node threshold.

From Eq. (8) we have that for $d > 0$, the system has at most a couple of interior zero points, $0 < z_{-1}, z_0 < 1$, in addition to the trivial ones: $z = 0, 1$. The previous double root $z = 0$ is now disentangled to two distinct single roots, $z = 0, z_{-1}$. The Z-side interior root, $\mathrm{P_Z}(d)$: $z = z_0$, is an asymptotically stable equilibrium corresponding to P for $d = 0$. The Y-side interior root, $\mathrm{P_Y}(d)$: $z = z_{-1}$, is an unstable equilibrium that separates the state space into two disjoint intervals: the basins of attraction respectively to node Y ($z = 0$) and $\mathrm{P_Z}(d)$. As the complexity cost $d$ increases from zero, $P_Y(d)$ emerges from the Y corner and appears in the state space, and then $P_Y(d)$ and $P_Z(d)$ move near each other, ultimately leading to a collision and vanishing at a global maximum point of $F_Z - F_Y$, which is a supercritical saddle-node bifurcation, as shown in Fig. 3.

To analyze the robustness of this equilibrium against complexity costs, we derive the critical threshold for $d$. By substituting $v := (1 - z)^{N-1}$ into the equilibrium condition $F_Z - F_Y = 0$ in Eq. (7), we obtain a quadratic equation with respect to $v$:

$$(b - c)v^2 - bv + c + d = 0. \tag{12}$$

The existence of real roots for $v$ (and thus for the equilibrium frequency $z$) requires the discriminant to be non-negative. This condition yields the saddle-node bifurcation point $d^*$, which represents the upper limit of the cost for the equilibrium to exist:

$$d^* = \frac{(b - 2c)^2}{4(b - c)}. \tag{13}$$

If the complexity cost exceeds this threshold ($d > d^*$), the stable mixed equilibrium disappears.

### 3.3 Invasion of unconditional cooperators

We here investigate how stable the established equilibrium P of integrated reciprocators (Z) and unconditional defectors (Y) is in the presence of unconditional cooperators (X).

**Theorem.** *P is a globally asymptotically stable rest point on the state space* $S_3$.

*Proof.* The image of X-players is fixed as Bad from the assessment rule given in Subsection 2.3. Becauase of this image, X-players will be treated as well as Y-players by Z-players. Regarding the expected payoff, X-players are disadvantageous to Y-players, as in the Prisoner's Dilemma game. This thus results in not only rare X-mutants being unable to invade P but also X being strictly dominated by Y in the interior of the X-Y-Z state space.

Consequently, all trajectories in the interior of the triangular state space converge to the edge between Z and Y, making the boundary equilibrium P the global attractor even for the replicator dynamics of the three strategies. ■

Overall, given the clear distinction between integrated reciprocators (Z) and unconditional defectors (Y), the population's evolutionary fate is the coexistence of these two types, with the eventual extinction of unconditional cooperators (X).

### 3.4 Non-zero complexity costs and invasion of conditional strategists

We then investigate how robust the stable equilibrium $P_Z(d)$ is against invasion of rare mutants with other different conditional strategies, W, which conditionally choose with whom to cooperate. Needless to say, there is an infinite number of variations of conditional strategies. However, we will show below that, given the strictness of the image assessment, it is possible to analyze the invasion potential of any such strategy generally.

In fact, under the dual-verification assessment rule, any mutant strategy that deviates from the integrated logic (e.g., a "pure downstream reciprocator" who refuses to pay-it-forward) is immediately assessed as "Bad" by resident Z-players. Under these conditions, the clear consequence is that resident Z-players will treat mutant W-players in the same way as Y-players (defecting against them). Since Y-players also defect, the gross benefit for W-players cannot exceed that of Y-players.

Crucially, however, unlike Y-players who are cognitively blind, conditional W-mutants incur a strictly positive complexity cost ($d > 0$) to process information. This renders the mutant strategy strictly dominated by the unconditional defector strategy Y. Since Y-players are already present in the mixed equilibrium, any such mutant is inevitably driven to extinction. Thus, counterintuitively, the existence of complexity costs "hardens" the stability of the polymorphism, actively protecting it against invasion by alternative conditional strategies.

# 4 Discussion

Our extension of integrated reciprocity to more general *N*-player games reveals an unexpected result: Stable behavioral diversity, not uniform cooperation, emerges as the evolutionary outcome. Rather than eliminating defectors as most indirect reciprocity models attempt (Okada, 2020; Santos et al., 2021), our framework maintains a mixed equilibrium of integrated reciprocators and unconditional defectors that persists without additional mechanisms like repetition or spatial structure.

This challenges the conventional view that defection represents a "problem" to be solved. In our model, defectors serve a structural role: They prevent invasion by unconditional cooperators (second-order freeloaders) who would otherwise destabilize the reciprocal system. The resulting polymorphism is asymptotically stable precisely because it includes defection, which is the very element traditional models seek to eliminate.

The robustness of this diversity is striking. The mixed equilibrium emerges endogenously through negative frequency dependence rather than from exogenous errors or mutations. In the model, negative frequency dependence arises from structural tension between unconditional forwarding and conditional rewarding: Forwarding aids defectors more when they are rare (providing them unreciprocated benefits), while rewarding aids reciprocators more when they are rare (concentrating benefits among fewer Good players).

Moreover, introducing complexity costs strengthens rather than weakens this stability by creating an additional barrier against alternative conditional strategies. This suggests that the cognitive burden of tracking both emotional and reputational information may actually protect behavioral diversity rather than undermine it.

We do not claim to have "solved" the cooperation problem—indeed, our equilibrium explicitly includes substantial defection that increases with group size. Instead, we demonstrate that strategic diversity itself, maintained through the integration of emotional and strategic reciprocity channels, can be evolutionarily adaptive. This reframes the question from "how to achieve full cooperation" to "why does behavioral diversity persist?"

The model generates specific testable predictions regarding group size. While conceptually aligned with Simpson et al.'s (2018) finding that gratitude and reputation function as dual mechanisms, our model uniquely addresses the dynamics of larger groups. Whereas their experiments focused on fixed small groups, our model predicts that the frequency of integrated reciprocators decreases as $N$ increases, suggesting that even combined mechanisms struggle with scale. Furthermore, our analytical derivation of the critical cost $d^*$ in Eq. (13) highlights a fundamental cognitive constraint. If the complexity cost scales linearly with group size (e.g., $d \propto N$), as is plausible when tracking multiple individuals, the cost will eventually exceed the threshold $d^*$. This leads to the collapse of cooperation in very large groups, a prediction that aligns with empirical findings (Shinada and Yamagishi, 2008). This divergence provides a critical research question: Does the behavioral switching pattern we predict become less frequent or merely less effective as $N$ increases? Moreover, the critical threshold $b/c = 2$ for stable polymorphism provides another test using Simpson et al.'s (2018) experimental paradigm.

It is also important to note the scope of our interactions relative to direct reciprocity. Our model assumes a large, well-mixed population where the probability of re-encountering the same individual is negligible, rendering direct reciprocity evolutionarily neutral. In contrast, in finite or structured populations, direct reciprocity plays a fundamental role and can often synergize with or supersede indirect mechanisms (Schmid et al., 2021). Recent work suggests that in such finite settings, direct reciprocity may even dominate upstream reciprocity when they compete (Pal et al., 2024). Therefore, our findings regarding the integration of upstream and downstream reciprocity are particularly relevant for large-scale societies where direct reinforcement loops are structurally absent.

While our theoretical insights emerge from a simplified model that assumes public assessment, extending them to more realistic settings raises important questions. Real-world reputation systems involve private, noisy assessment where individuals may disagree about others' reputations (Uchida, 2010; Okada et al., 2017, Hilbe et al., 2018, Uchida et al., 2018; Fujimoto and Ohtsuki, 2023). The integrated strategy's pay-it-forward component operates independently of reputation assessment, potentially providing robustness against assessment disagreement. However, formally analyzing integrated reciprocity under private assessment requires extending our framework substantially and remains an important direction for future work.

The extensive literature on indirect reciprocity has primarily focused on assessment rules—from Ohtsuki and Iwasa's (2004, 2006) identification of the "leading eight" norms to recent expansions into higher-order rules (Santos et al., 2018). This "tournament of assessment rules" consistently seeks norms that eliminate defectors entirely. However, Murase and Hilbe (2024) demonstrated that even sophisticated third-order assessment rules struggle to achieve full cooperation in well-mixed populations, suggesting fundamental limits to the assessment-focused approach.

Our work diverges from the traditional paradigm that seeks ever-more-complex assessment rules to eliminate defectors (Ohtsuki and Iwasa, 2006; Santos et al., 2018). Instead, we demonstrate that modifying action rules—specifically by integrating upstream and downstream reciprocity—can actively maintain strategic diversity. By embracing the coexistence of cooperators and defectors, our model offers an evolutionary pathway well-suited to the realities of unstructured populations, where simple assessment rules are often the norm. This shift suggests that exploring the space of action rules, which has been largely neglected in favor of assessment complexity, is a critical frontier for understanding human cooperation.

Furthermore, the stability of our integrated strategy draws an interesting parallel to the literature on tag-based cooperation (Riolo et al., 2001; Hammond and Axelrod, 2006). In those models, individuals cooperate based on observable traits (tags), forming exclusive cooperative clusters similar to the Z-player clusters in our equilibrium. However, a known vulnerability of pure tag models is the invasion of "cheaters with the right tag"—mutants who possess the cooperative tag but refuse to pay the cost of helping (Roberts and Sherratt, 2002). Our model addresses this vulnerability by grounding the "tag" (reputation) in dynamic action rather than static traits. Through the dual-verification mechanism, the "Good" label becomes a status continuously earned by adhering to both upstream and downstream norms, effectively filtering out second-order free-riders who would otherwise mimic a static cooperative tag.

We believe that the theoretical framework established in this study provides a foundational step toward bridging the gap between reputation-based and tag-based evolution of cooperation. By demonstrating how dynamic reputations can function with the robustness of observable tags, our findings offer a concrete basis for future research aiming to integrate these distinct mechanisms into a unified understanding of social dilemmas.

## Conflict of Interest

The authors declare that the research was conducted in the absence of any commercial or financial relationships that could be construed as a potential conflict of interest.

## Author Contributions

T.S. initiated the project and performed writing, formal analysis, investigation, and visualization. All authors contributed to conceptualization, methodology, review and editing. All authors have read and agreed to the published version of the manuscript.

## Funding

This work was supported by JSPS KAKENHI, Grant Numbers 23K05943(TS, YN), 23K21017(IO, HY), 21KK0027(IO, HY), 23K25160(HY, IO), 25K21907(IO).

**Figures and figure captions**

**Figure 1**

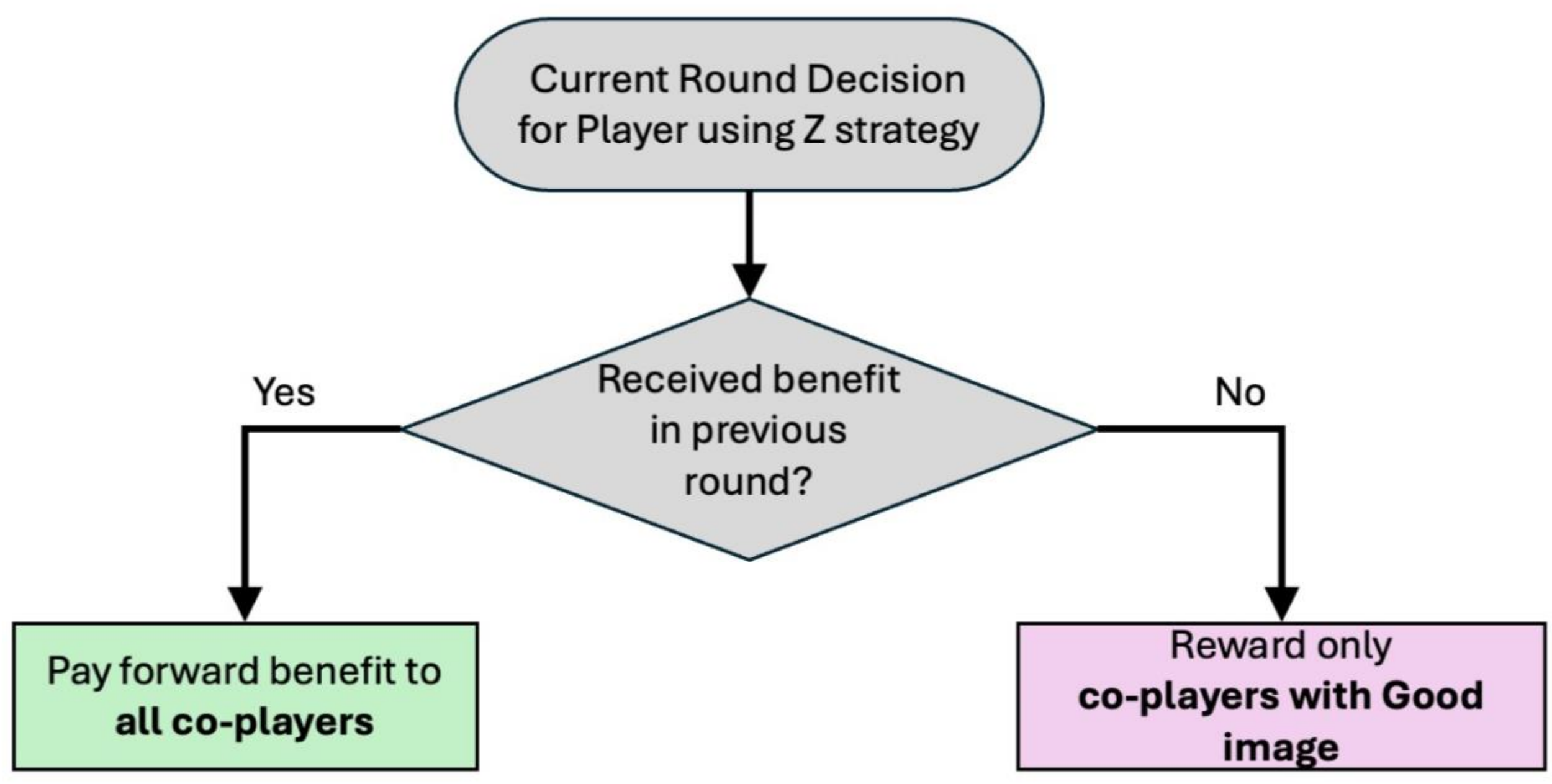

Figure 1: Decision logic and state-action map for the integrated (Z) strategy. A player helps if (i) she was helped in the previous round or (ii) her current co-players have a Good image.

**Figure 2**

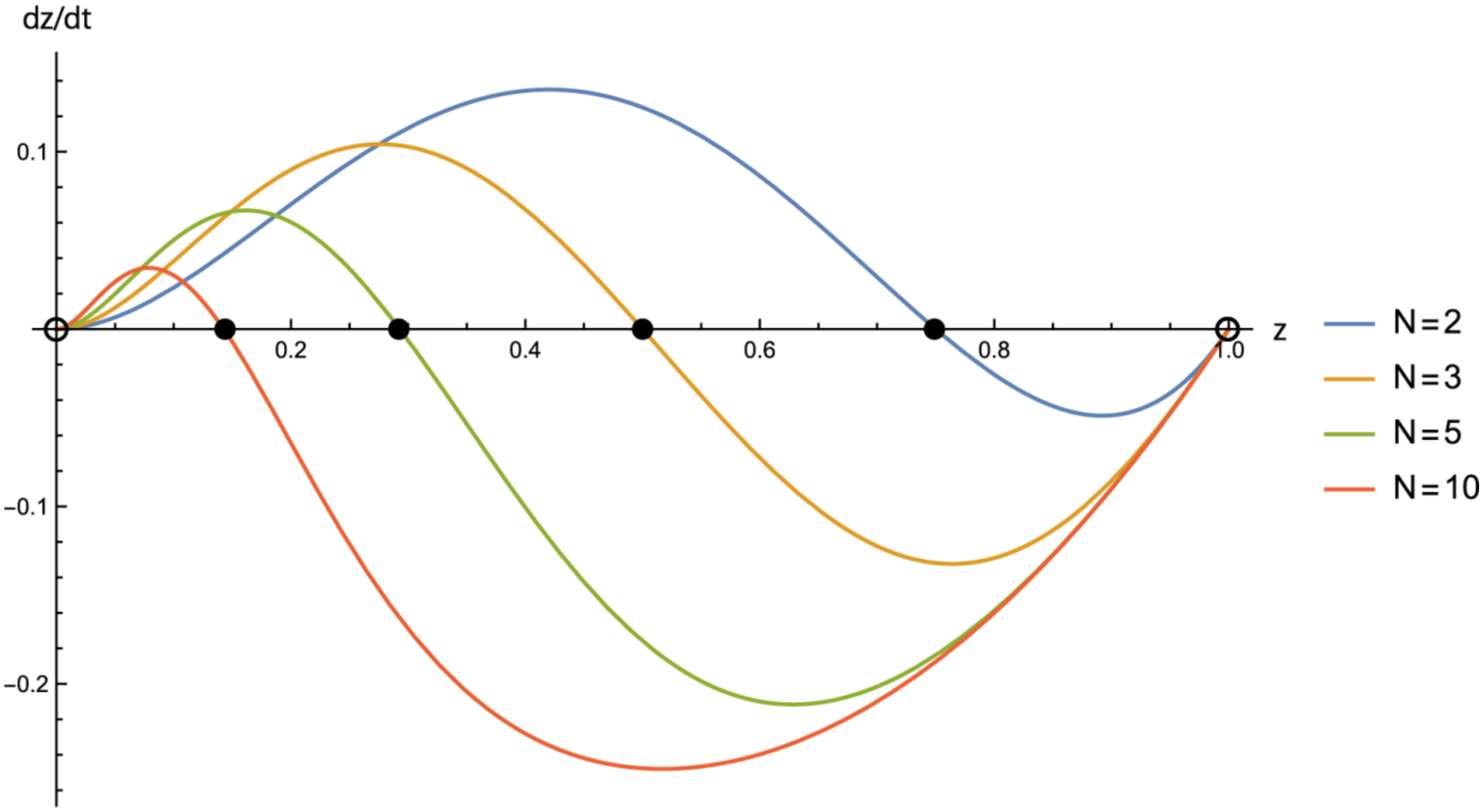

Figure 2: Selection gradient $dz/dt$ (gradient of replicator dynamics on Y-Z edge) for group sizes $N = 2$ (blue), 3 (orange), 5 (green), 10 (red). Parameters are: $b = 5$, $c = 1$, and $d = 0$. Filled circles denote asymptotically stable equilibria; open circles denote unstable equilibria.

**Figure 3**

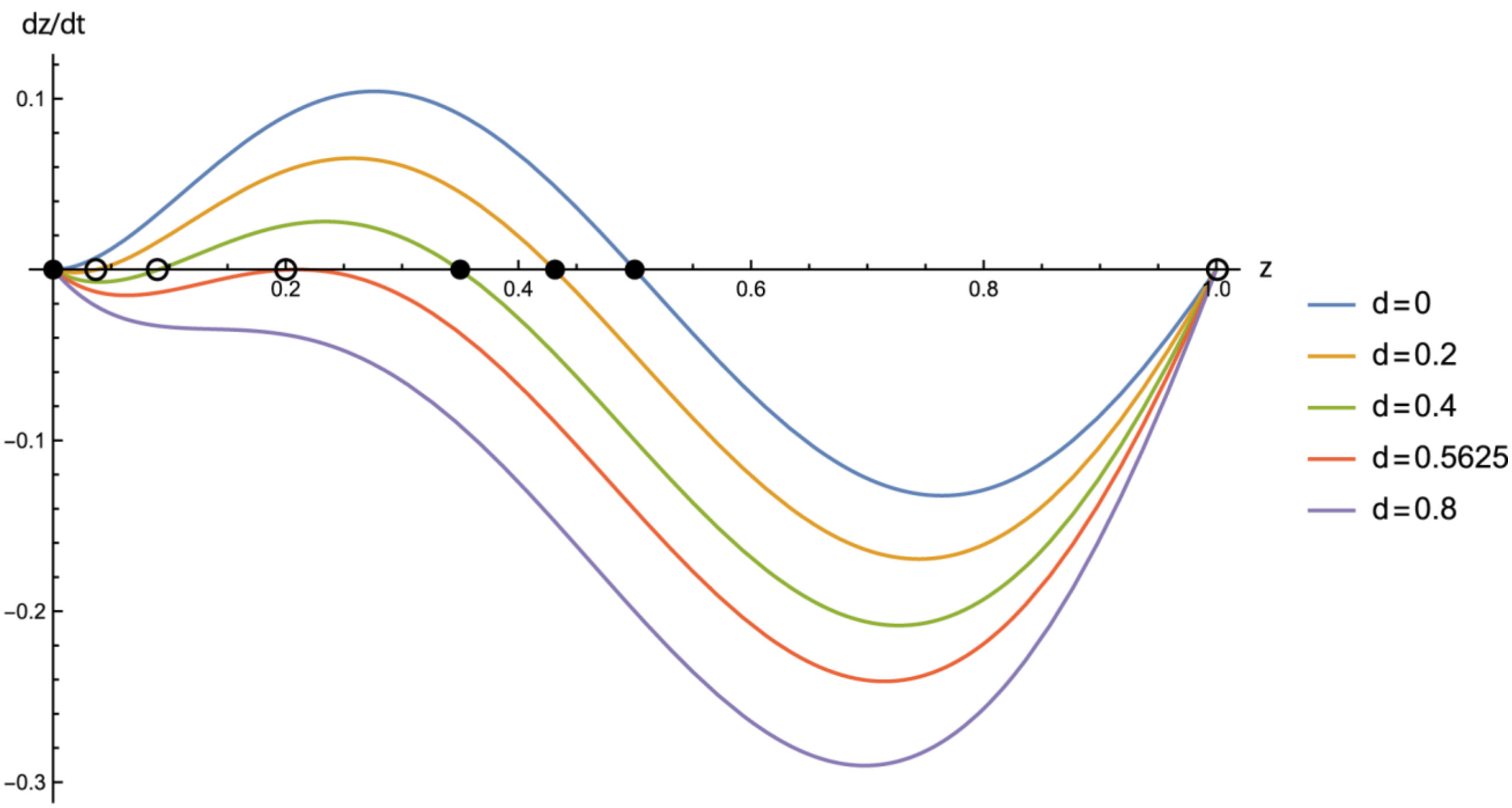

Figure 3: Selection gradient $dz/dt$ (gradient of replicator dynamics on Y-Z edge) for complexity costs $d = 0$ (blue), 0.2 (orange), 0.4 (green), 0.5625 (red), 0.8 (purple). Parameters are: $b = 5$, $c = 1$, and $N = 3$. Filled circles denote asymptotically stable equilibria; open circles denote unstable equilibria. Notably, the red curve ($d = 0.5625$) corresponds to the critical saddle-node bifurcation point $d^*$, where the interior stable and unstable equilibria collide and vanish.

**Table**

**Table 1**

| | Player using Z strategy in $N$-player group<br>($N_Z$ : number of players using Z strategy) | |
| --- | --- | --- |
| | pays forward benefit $b$<br>shared among<br>**all co-players** | gives reward $b$<br>shared among<br>**only Good co-players** |
| Co-player using Z strategy receives | $\frac{b}{N-1}$ | $\frac{b}{N_Z-1}$ |
| Co-player using Y strategy receives | $\frac{b}{N-1}$ | 0 |

Table 1: Benefits received by different strategy types when paired with an integrated reciprocator (Z) donor in $N$-player groups